\documentclass[conference]{IEEEtran}
\IEEEoverridecommandlockouts
\usepackage{cite}
\usepackage{amsmath,amssymb,amsfonts}
\usepackage{algorithmic}
\usepackage{balance}
\usepackage{graphicx}
\usepackage{textcomp}
\usepackage{xcolor}
\usepackage{hyperref}

\def\BibTeX{{\rm B\kern-.05em{\sc i\kern-.025em b}\kern-.08em
    T\kern-.1667em\lower.7ex\hbox{E}\kern-.125emX}}
\begin{document}

\title{Cloud Revolution: Tracing the Origins and\\Rise of Cloud Computing\\
}

\author{

\IEEEauthorblockN{Deepa Gurung*}
\IEEEauthorblockA{Business Administration \\
Joongbu University, Seoul, South Korea \\
email: grgdips110@gmail.com}
\and
\IEEEauthorblockN{S M Zia Ur Rashid*}
\IEEEauthorblockA{Department of Electrical and\\ Computer Engineering \\
The University of Tulsa, Tulsa, USA \\
email: ziaur-rashid@utulsa.edu}
\and
\IEEEauthorblockN{Zain ul Abdeen*}
\IEEEauthorblockA{Bradley Department of Electrical \\and Computer Engineering \\
Virginia Tech, Blacksburg, USA \\
email: zabdeen@vt.edu}
\and
\IEEEauthorblockN{Suman Rath}
\IEEEauthorblockA{Department of Electrical and\\ Computer Engineering \\
The University of Tulsa, Tulsa, USA \\
email: suman-rath@utulsa.edu}
}


\maketitle

\begin{abstract}
The history behind the development of cloud computing is more than several decades of technological progress in the fields of virtualization, distributed systems, and high-speed networking, but its current application is much broader than the underlying technologies that made it possible. This paper reexamines the historical evolution of the field, including the initial ideas of resource sharing and utility-based computing approaches and the development of hyperscale data centers and modern globally federated cloud ecosystems. We also analyze the technological and economic forces and point to the way cloud platforms altered the organizational computing habits, decreasing the entrance-level to the data-intensive and computation-heavy apps. The study also takes into account the ongoing limitations which have come with the large-scale adoption of clouds which include exposure to security due to the weaknesses in configuration, particular establishment regulations, and structural reliance on the single vendors. Lastly, we address some of the new trends that are transforming the cloud environment, including the convergence of edge and cloud infrastructure, the increased prominence of AI-optimised architectures and the initial adoption of quantum computing services. Collectively, the developments above describe an emerging but quickly changing paradigm with its future direction being determined by a strike of balancing between scalability, openness, and trust.

\end{abstract}


\section{Introduction} 
In today’s tech-driven modern era, cloud computing serves as the backbone that enables global accessibility to advanced capabilities in artificial intelligence (AI), quantum technologies (QT), and edge computing \cite{armbrust2010view, buyya2009cloud}. Both AI and quantum technologies demand sophisticated computational infrastructure and massive processing capabilities that traditionally only well-resourced organizations such as IBM for quantum computing and Google for large-scale AI could afford \cite{preskill2018quantum, chollet2017deep}. Cloud computing bridges this accessibility gap by delivering on-demand access to high performance computational resources \cite{marston2011cloud}. Consequently, students, researchers, small enterprises, and other stakeholders are enabled to use sophisticated tools such as AI and QT without having to invest a lot of capital.

The pay-per-use economic model associated with cloud computing has fundamentally altered how organizations provision and manage IT infrastructure \cite{mell2011nist}. Rather than purchasing and maintaining servers, storage arrays, and networking equipment, customers allocate virtual resources elastically and are billed according to consumption. Cloud providers have also become innovators in various industries such as education, finance, healthcare and manufacturing, allowing start-ups to grow rapidly without bearing the expensive infrastructure expenses which would be incurred under conventional resource provision models of infrastructural ownership \cite{sultan2010cloud}.
At the same time, the public cloud market has consolidated around Amazon Web Services (AWS), Microsoft Azure, and Google Cloud Platform (GCP). As of Q3 2025, the three together account for ~62\% of global cloud infrastructure services; individually AWS ~29–30\%, Azure ~20\%, Google Cloud ~13\% \cite{synergy2025share, statista2025share}.
It is expected that in 2025 the global market of cloud computing will experience a growth up to \$0.86 trillion and in 2030 will grow to an even greater number of 2.26 trillion, which represents an increase of 21.20\% per year and indicates its role in the macroeconomy \cite{gartner2018magic}. Although scalability and cost-effectiveness benefits have influenced this massive adoption, it also creates significant strategic trade-offs. Organizations must weigh these benefits against substantial risks, including security threats, regulatory complexity, performance bottlenecks, and vendor lock-in to make informed decision and strategic planning \cite{fernandes2014security, zhang2010cloud}.

{Given these factors, we aim to provide a stringent analysis of the development, functional capabilities, and challenges that relate to cloud computing. The history begins with mainframe time-sharing systems in 1960s then the launch of AWS in the year 2006 up to the current hyperscale infrastructures. The main advantages that drive the mass adoption, such as scalability, cost effectiveness, and availability are evaluated, and persistent issues of security, compliance, and vendor independence are critically evaluated. Moreover, emerging paradigms such as edge computing, serverless architectures, and quantum-cloud integration are analyzed, as these innovations are expected to significantly influence the trajectory of cloud computing over the next decade.}

{We structure the remainder of this manuscript as follows. Section \ref{historical} reviews the historical evolution of cloud computing from early time-sharing to contemporary hyperscale infrastructures and service models. Section \ref{benefits_of_CC} addresses the advantages of cloud computing on a scalability, cost, availability and performance. Section \ref{challenge} examines some of the significant issues and constraints, such as security, interoperability, compliance, and cost management. Section \ref{sec5} explores emerging technologies and future trends at the intersection of cloud, edge, AI, IoT, big data, and quantum computing. Finally, in Section \ref{conc}, we conclude the paper and outline directions for future research and practice.}





\section{Historical Evolution of Cloud Computing}\label{historical}
\begin{figure*}[ht]
  \centering
  \includegraphics[width=\textwidth, trim=0.3cm 5cm 0.3cm 5cm, clip]{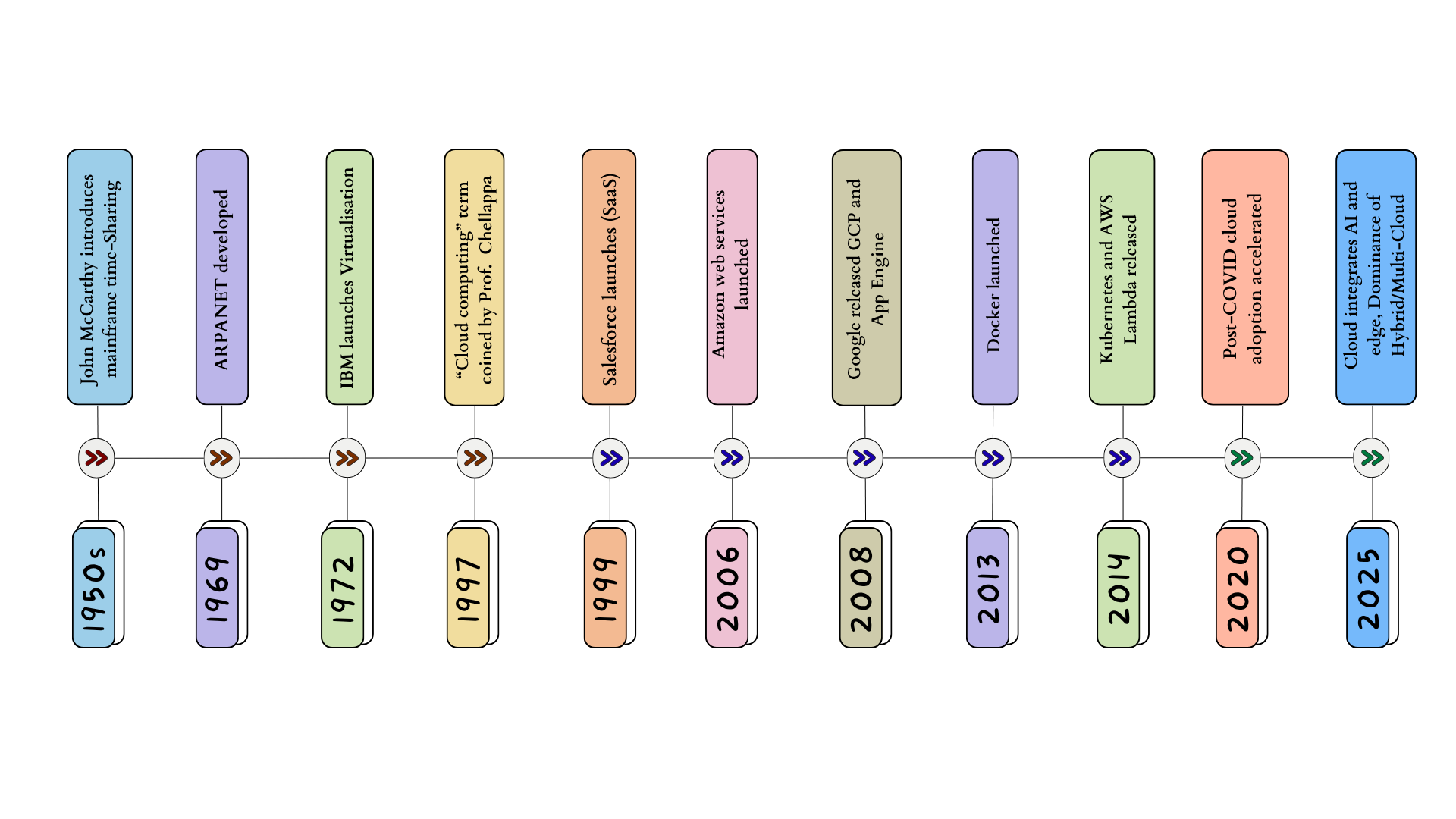}
  \caption{The Genesis of the Cloud Computing}
  \label{fig:history}
\end{figure*}
\subsection{Early Foundations}
As shown in Fig. \ref{fig:history}, the conceptual precursors of cloud computing could be dated back to the 1950s and 1960s and are associated with mainframe computing during the time when prohibitive costs of hardware led to the development of innovative strategies for sharing resources. Time-sharing systems emerged, enabling multiple users to concurrently access a central computer, establishing early principles of shared resources and cost efficiency through economies of scale \cite{licklider1968computer, parkhill1966challenge}. These architectures laid the basis of the principles of multiplexing, isolation, and effective use of resources that would later be experienced in the virtualized and cloud environment. The concept of computing as a service, with processing and storage being consumed on-demand instead of owned, was also a precursor of time-sharing.

In the 1970s and 1980s, distributed computing systems were developing, in which interconnected computers would work together to perform complex tasks, thereby providing important foundation for decentralised resource management \cite{tanenbaum2007distributed}. The commercialization of the internet during the 1980s further accelerated these developments, enabling remote network-based access to geographically distributed computing resources \cite{leiner2009brief}. At the same time, the virtualization technologies invented by IBM enabled the use of several operating systems on one physical machine, which provided new possibilities of flexibilities and optimization of the resources \cite{goldberg1974survey, rosenblum2005virtual}.
All these developments formed the technological foundations upon which contemporary cloud platforms are constructed.

\subsection{Key Milestones}

The 1990s introduced utility computing, conceptualizing computing resources as metered services analogous to electricity. Computer scientist John McCarthy famously proposed in 1961 that \textit{computation may someday be organized as a public utility}, a vision that began materializing as companies like IBM and Sun Microsystems experimented with on-demand computing resources \cite{rappa2004business, buyya2009cloud}.

The late 1990s marked the emergence of SaaS as a viable delivery model. Founded in 1999, Salesforce proved that traditional complex enterprise applications like customer relationship management (CRM) system could indeed be fulfilled completely through a web interface on a pay-per-use model \cite{chong2006software}, thereby proving both the technical conceptualization and business viability of internet-delivered software.

The watershed moment arrived in 2006 when AWS, introducing Elastic Compute Cloud (EC2) and Simple Storage Service (S3) as foundational infrastructure services \cite{vaquero2008break}. AWS revolutionized the landscape by providing scalable, elastic, pay-as-you-go infrastructure accessible to developers and businesses of all sizes. This operationalized Infrastructure-as-a-Service (IaaS) at a global scale and catalyzed cloud-native application design \cite{armbrust2010view}. In response, major technology companies entered the market. In 2010, Microsoft launched Azure which at first focused on PaaS and later expanded into an overall hybrid-cloud platform \cite{copeland2010microsoft}. The internal infrastructure of Google was redesigned into GCP with its advantages of massive data processing and machine learning \cite{bisong2019google}. The rivalry between these providers stimulated the innovation of elasticity, storage, networking, and managed services \cite{gartner2018magic}.

Businesses were beginning to employ hybrid and multi-cloud strategies in the middle of the 2010s to have more control, flexibility, and resilience. Hybrid deployments kept sensitive workloads on private infrastructure while using public clouds for burst capacity \cite{opara2016multiclouds}. Multi-cloud strategies were used to mitigate vendor lock-in or align workloads with providers offering specific advantages. In parallel, the rise of mobile, IoT, and real-time applications highlighted the limitations of centralized architectures, motivating the development of edge computing, which places computation closer to data sources and end users \cite{satyanarayanan2017emergence, pham2020survey}.

\subsection{Data Center Evolution}

Cloud infrastructure relies on physical, hyperscale data centers which are massive facilities optimized for scale, operational automation, and energy efficiency \cite{barroso2013datacenter}. To achieve fault isolation, redundancy, and low-latency, these centers are geographically distributed into regions and availability zones, a strategy driven by massive investment from companies like Amazon, Microsoft, and Google \cite{barroso2013datacenter, greenberg2008cost}. Virtualization has contributed to the development of such infrastructure by creating the ability to consolidate workloads \cite{greenberg2008cost}, and later using container orchestration software like Kubernetes that standardized application deployment and scaling \cite{kreutz2015software}. Concurrently, Software-Defined Networking (SDN) separated control of the network, allowing flexible traffic management for multi-tenant workloads \cite{aws2020architecture}.

\subsection{Cloud Computing Architecture and Models}
\begin{figure}[htb]
    \centering
    \includegraphics[width=\linewidth, trim= 16cm 1.5cm 10.5cm 1.5cm, clip]{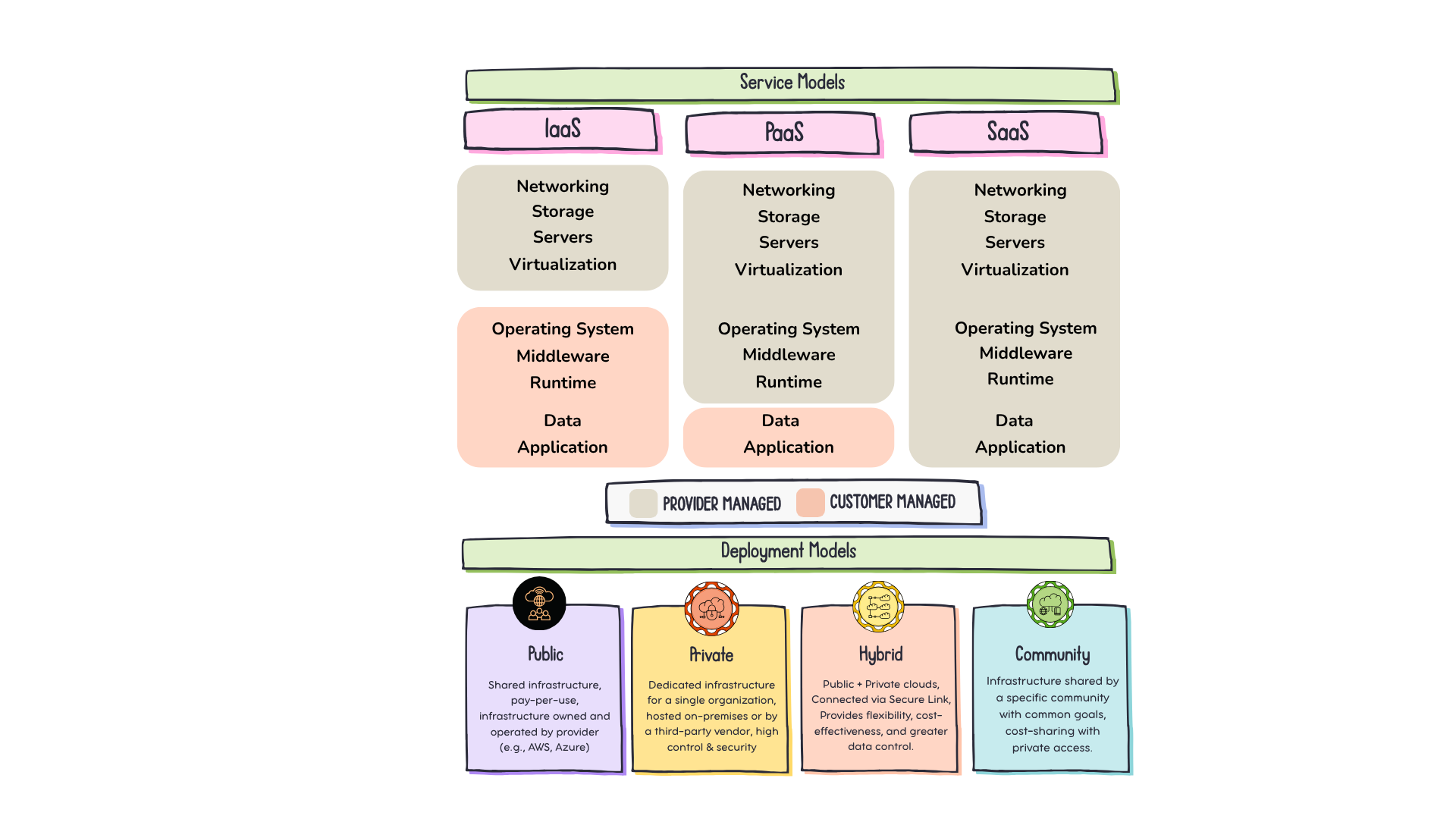}
    \caption{Deconstructing the Cloud: Service vs. Deployment}
    \label{fig:models}
\end{figure}

Cloud computing is defined by service and deployment models that frame the trade-offs in control, abstraction, and responsibility between provider and consumer. As illustrated in Figure~\ref{fig:models}, it delivers services through three primary service models and four deployment models that define how resources are provisioned and accessed \cite{mell2011nist}.

\subsubsection{Service Models} The service models are a hierarchical framework of abstraction. Infrastructure-as-a-service (IaaS) includes hardware facilities that have been provided in a virtualised form (e.g., Amazon EC2, Google Compute Engine), and in this case, the customer takes control of operating system and application management. PaaS layers over the infrastructure, and provides a managed environment (e.g. Heroku, Azure App Service) in which the developer can deploy and scale apps \cite{fehling2014cloud}. SaaS offers complete, provider-managed applications (e.g., Salesforce, Microsoft 365) to end-users over the network \cite{chong2006software}.

\subsubsection{Deployment Models} Public clouds offer scalable, shared, multi-tenant infrastructure over the internet \cite{buyya2009cloud}. On the other hand, private clouds provide dedicated resources to the individual organization thus can afford more control and compliance at high costs \cite{marston2011cloud}. Hybrid cloud model seeks a middle ground, combining both public and private environments to allow workload portability based on the need to maintain security or need to scale up \cite{cavalcante2016hybrid}. Lastly, multi-cloud architectures use multiple public providers to reduce dependency but add significant operational complexity in orchestration and monitoring \cite{opara2016multiclouds}.

\section{Benefits of Cloud Computing}\label{benefits_of_CC}
Cloud computing offers distinct advantages over on-premises infrastructure, primarily by reconfiguring IT economics and resource management. In this section, we analyze three central benefits: scalability and elasticity, economic advantages, accessibility and performance.

\subsection{Scalability and Elasticity}
One thing that makes a cloud unique is its ability to automatically increase or decrease resources based on demand. Providers enable both vertical and horizontal scaling to accommodate fluctuating workloads without requiring substantial architectural redesign ~\cite{herbst2013elasticity, lorido2014auto, lehrig2015cloud}. Vertical scaling (scale-up) increases the computational or memory resources available to a virtual instance, which is particularly useful for database workloads and stateful applications that cannot easily be partitioned. By contrast, horizontal scalability (scale-out) distributes load by adding or removing replicas of stateless components, such as web servers or microservices, in response to demand \cite{lorido2014auto}. Elasticity automates this scaling based on real-time metrics. Strong auto-scaling services are available via the cloud providers, which keep an eye on the application measurements and allocate or destroy resources on-demand as necessary \cite{herbst2013elasticity}. For example, an e-commerce site can automatically add servers for a holiday sale or a streaming service like Netflix can handle viewer surges for a content release. The result of this elasticity is cost efficiency achieved through overprovisioning elimination, enhanced performance during peak traffic, agility in the business to respond quickly to market changes, and ability to withstand resilience of critical services \cite{armbrust2010view}.

\subsection{Economic Advantages}
Cloud computing offers significant economic benefits by shifting infrastructure acquisition and maintenance from capital expenditure (CapEx) to operational expenditure (OpEx) \cite{marston2011cloud, iyer2013cloud}. Instead of purchasing and depreciating hardware, organizations pay for compute, storage, and network resources based on usage. The pay-per-use pricing structure allows organizations to pay only for resources they use and avoid overprovisioning for peak demand \cite{al2012cloud}. During off-peak hours, systems can scale down, proportionally reducing expenses. As an example, a small e-commerce business can scale its resources on GCP in case of a significant sale and then downscale and only pay the cost of the resources that were actively used. Secondly, hyperscale providers are highly energy efficient due to the economies of scale, and this reduces the per-customer energy usage \cite{barroso2013datacenter}. This economic benefit is added with a decreased operational load since these providers take care of all infrastructure maintenance, patch management and security functions \cite{sultan2010cloud}. Additionally, cost monitoring and optimization services like AWS Cost Explorer and similar services on other platforms help the company to know which resources are idle or unutilized, adjust reservations, and trade-offs between types of instances or storage classes \cite{khajeh2011resource}.

\subsection{Accessibility and  Performance}

Cloud computing fundamentally changes user access by delivering services over the internet, making applications available from any device, anywhere. Professionals work remotely using platforms like Google Workspace or Microsoft 365, accessing documents and collaborative tools from multiple devices. Students access educational resources remotely via platforms like Google Classroom and Coursera.

Cloud platforms enable geographically dispersed users to collaborate simultaneously on shared documents or projects. The combination of features like the version control and shared working environment and real-time editing promotes a smooth process of teamwork and helps to avoid confusion related to multiple copies of a file or outdated versions. By centralizing data in the cloud, all authorized users access the most current information which reduces data silos, enhances transparency, and greatly simplifies backup and recovery procedures \cite{sultan2010cloud}.

From technical standpoint, cloud computing substantially improves system performance and reliability, establishing it as critical infrastructure. Cloud providers also have state-of-the-art data centres that have modernized hardware, streamlined storage offerings and high-speed networking \cite{greenberg2008cost}. Providers use Content Delivery Networks (CDNs) to serve large user base and minimize latency and provide services such as Netflix and YouTube to stream high-definition video to millions of users simultaneously without buffering. Cloud architectures are additionally designed for high availability. Providers use automated failover, redundancy, and geographic distribution to back service-level agreements (SLAs) that guarantee high uptime (e.g., 99.9\%). Cloud computing also simplifies the disaster recovery and business continuity planning through automated backup and recovery provisions. For instance, a bank using AWS Disaster Recovery can replicate its systems in geographically distant locations, which in turn will guarantee business continuity in case of cyber-attacks or natural disasters. Finally, advanced load-balancing mechanisms distribute workloads evenly across servers, maintaining application responsiveness and preventing bottlenecks during traffic peaks \cite{greenberg2008cost}.

\section{Challenges and Limitations}\label{challenge}
Cloud computing as a technology has numerous benefits. Despite this, there can be several significant challenges that organizations may face and need to account for to ensure effective and strategic adoption of the cloud revolution. The following section provides a comprehensive analysis of a few such limitations that form primary barriers to the adoption of cloud computing.

\subsection{Security and Privacy Concerns}
Security is one of the most important problems that must be addressed in cloud computing. It is of such paramount importance that organizations must be careful when entrusting sensitive data to external cloud providers \cite{fernandes2014security}.
Cloud computing environments may face several threats and potential data breaches stemming from one or more of the following reasons: misconfigured storage or weak access controls, insider misuse of privileged accounts, and vulnerabilities in application programming interfaces (APIs). Vulnerabilities in APIs often serve as attack vectors. The case of the 2019 Capital One breach, which was caused by a poorly set up AWS firewall that leaked more than one hundred million customer records. This is a good example of how one configuration mistake can lead to a massive data leakage \cite{pearson2013privacy}.

Another aspect to be concerned about here is privacy.
These concerns become more pronounced in multi-tenant clouds, where different customers typically share the same physical resources and have to rely on isolation mechanisms they have no direct control over \cite{pearson2013privacy}. Cross-border data transfers are a notable concern here as different jurisdictions may have different privacy compliance regulations.
Hence, to  mitigate these risks, organizations must attempt several fortifications. Some of these are: strong encryptions for static and mobile (in-transit) data, rigorous access control and identity management, periodic/aperiodic security audits, and ensuring adherence to frameworks such as GDPR, HIPAA, and CCPA \cite{almorsy2016cloud}. However, in the end, it is also important to understand and accept that it can be difficult to assess and manage security compliance once data leave the home environment and flow to provider-side controls.

\subsection{Vendor Lock-In and Interoperability}
Vendor lock-in denotes a situation in which organizations become dependent on the proprietary tech from specific cloud providers. Being dependent on such technologies often means that customers will be bound by technological and financial constraints when attempting to move to alternative platforms making it heavily prohibitive. The proprietary tech from cloud providers typically includes APIs, data formats, and services that lack standardization across platforms. One of the major examples of this are applications built using AWS Lambda serverless functions. These applications cannot be easily migrated to another platform like Azure Functions without substantial code refactoring \cite{opara2016multiclouds}.

The problem of vendor lock-in is often elevated by data portability challenges. This is because, extracting large amounts of stored data from cloud incurs proportionally huge amounts in egress fees. Vendor lock-in is a significant strategic risk as it reduces organizations' negotiation leverage with providers and limits their options towards using best-of-breed services provided by different vendors. Mitigation of this condition is typically achieved by using multi-cloud strategies and containerization technologies like Docker and Kubernetes for portability. Implementing abstraction layers that separate applications from provider-specific services and prioritize open standards and APIs when selecting cloud services also helps to ameliorate the impact of lock-in \cite{opara2016multiclouds}.


\subsection{Compliance and Regulatory Challenges}
Companies using cloud computing are often dispersed geographically. These areas are typically governed by varying types and levels of data residency and privacy laws, complicating multi-region cloud deployment (especially when the cloud servers are located in a different nation-state). As an example, the European Union's General Data Protection Regulation (GDPR) sets prohibitively strict data-processing and storage requirements and heavy penalties in case of non-compliance. An excellent example of this paradigm would be the healthcare organisations, which are required to be compliant with the HIPAA regulations on patient health information and cannot send such information to cloud without any express approvals. The financial institutions also have similar strict regulatory framework. To alleviate these concerns, organizations must vet provider data protection standards and compliance certifications. Furthermore, robust encryption, maintain detailed audit trails, and conduct regular compliance assessments are important elements of a strong governance strategy \cite{pearson2013privacy}.

\subsection{Latency and Cost Management}
Although cloud computing as technology has many performance benefits, network latency remains a critical concern. Network latency can negatively impact real-time responsiveness of critical systems dependent on cloud services. Latency-sensitive applications including high-frequency trading, real-time gaming, and industrial control systems may experience unacceptable delays when processing occurs in distant data centers.
Applications dependent on the internet face significant service disruptions during network outages. In shared multi-tenant environments, the so-called noisy neighbor effect may slow down performance as a single tenant consumes an unequal portion of resources to the detriment of the rest. Simultaneously, cost management remains a critical challenge. Although the emergence of cloud computing has reduced costs, elastic scaling may result in several unforeseen expenses, which eventually makes budgeting very challenging \cite{al2012cloud}. Unpredicted expenses such as data egress fees, API call charges, and premium support rates can contribute significantly to the overall expenses. Cost overruns can also be caused by an over-provisioned resources or inefficient architectures. This problem can be alleviated by adequate monitoring, resource tagging, and efficient allocation strategies.

\section{Emerging Technologies and Future Trends}\label{sec5}
Cloud infrastructure provides the scalable power and data capabilities necessary for next-generation technologies like AI, IoT, edge and quantum computing. This section examines how the cloud services are integrated with emerging technologies and the trends likely to influence cloud evolution going forward.

\subsection{Artificial Intelligence and Machine Learning}
Cloud platforms already support comprehensive AI and machine-learning ecosystems, enabling enterprises to develop and implement intelligent applications without the need for specialized hardware investments \cite{armbrust2010view}. Managed services such as Vertex AI, SageMaker, and Azure ML allow users to run advanced tasks—image and speech recognition, natural-language understanding, and more—without running large clusters on-premises \cite{chollet2017deep}. These platforms provide automated workflows for model training and deployment, supported by cloud GPUs and TPUs necessary for large-scale neural networks.

Advanced AI capabilities, once the exclusive domain of well-resourced institutions, have been democratized by cloud platforms. This transition now permits small organizations and individual researchers to access powerful computational tools. Cloud-based AI applications cover diverse domains, including healthcare diagnostics, financial fraud detection, autonomous vehicles, and personalized recommendation systems. The fast expansion of this ecosystem reflects continued investment in specialized AI hardware and software optimized for distributed training and inference workloads \cite{zheng2019aiops}.

\subsection{Internet of Things and Edge Computing}
IoT ecosystems also depend heavily on cloud services to store, process, and analyze sensor data at scale. Cloud-based IoT platforms offer device management for securely connecting millions of devices, real-time data analytics for extracting actionable insights from streaming data, and integration with edge computing to reduce latency by processing data closer to sources before cloud transmission \cite{shi2016edge}.
Edge computing has become a new complementary paradigm of cloud computing, which puts computational systems at the network edge adjacent to the data sources and end-users as well as end-products \cite{satyanarayanan2017emergence}. This distributed architecture enables ultra-low latency processing critical for applications including autonomous vehicles, industrial automation, augmented reality, and smart city infrastructure. This intersection between edge and cloud computing is commonly referred to as the "edge-cloud continuum," which enables the allocation of workloads dynamically depending on the latency requirements, bandwidth limitations, and data privacy requirements \cite{pham2020survey}. Many smart-city projects already use cloud–edge systems to coordinate traffic, monitor air quality, manage energy use and improve public safety.

\subsection{Big Data Analytics and Quantum Computing}
Cloud platforms make it feasible for organizations to collect, store, and analyze data at petabyte scale by combining distributed computing frameworks, real-time processing engines, and data-lake architectures \cite{hashem2015rise}. Cloud-native services like Google BigQuery, Amazon Redshift, and Azure Synapse Analytics support large-scale queries and integrate predictive analytics and AI capabilities. Healthcare organizations, for example, use cloud analytics to examine patient records, detect disease outbreaks, and support drug-development pipelines \cite{hashem2015rise}.
Quantum computing, although still in the Noisy Intermediate-Scale Quantum (NISQ) stage, is increasingly mediated by cloud platforms \cite{preskill2018quantum}. Providers such as IBM Quantum Experience, AWS Braket, and Azure Quantum expose quantum processors and simulators via cloud interfaces, enabling experimentation without specialized hardware. Despite the modest capabilities of current hardware, these services provide researchers and organisations a practical way to explore quantum algorithms and to prepare for a future in which quantum advantage becomes attainable \cite{cerezo2021variational}.

\subsection{Future Directions}
There are several simultaneous trends that propose that cloud computing will remain to develop in various dimensions, as opposed to meeting at one point and location. The growing popularity of serverless and Function-as-a-Service (FaaS) paradigms is one of these directions \cite{jonas2019cloud, baldini2017serverless}. People who use these models deploy functions or event-driven components, and the underlying platform performs provisioning, scaling and placement. This decoupling is capable of simplifying the development of applications, as well as improving the efficiency of their resources, especially when it comes to bursty or intermittent workloads, but also increases the concern with observability, cold-start latency, and vendor dependency \cite{jonas2019cloud, baldini2017serverless}.

The other path is the development of confidential computing that aims at securing data not just when stored and moved but also processed. TEEs, including Intel SGX, AMD SEV, and ARM TrustZone, make it possible to compute on encrypted or shielded data to minimize the vulnerability of sensitive workloads to potentially compromised hypervisors or host operating systems \cite{fernandes2014security}. Confidential-computing services might transform trust assumptions and regulatory policies in multi-tenant settings due to the rollout of cloud providers offering such services.

The concept of sustainability will probably continue to be a prevailing design factor. The further proliferation of cloud and artificial intelligence workloads increases the attention to the issues of energy consumption and carbon emission \cite{koomey2011growth}. In response, major providers are aggressively pursuing efficiency through custom hardware, advanced cooling systems, and energy-aware workload placement \cite{barroso2013datacenter, gao2014machine}. Regulatory and market forces also encourage guiding and open reporting on environmental indicators and incorporating sustainability factors into working placement and scheduling choices.

Lastly, networking developments, particularly 6G technologies, could propose ultra-low latencies and higher bandwidths, enabling convergence in the performance of edge and cloud resources \cite{pham2020survey}. Together with AI, IoT, and potential quantum applications, these advances suggest that next generation of cloud infrastructures will become significantly more heterogeneous, distributed, and specialized compared to current paradigms and will involve a wide range of compute substrates, hierarchies, and more advanced trust models.

\section{Conclusion}\label{conc}

Cloud computing has transformed how the modern world accesses and uses IT resources.
It now serves as a foundational layer for emerging technologies, including artificial intelligence, IoT, and quantum computing.
It has also helped us shift from a range of locally controlled systems to a truly globally distributed utility model.
Despite its massive utility, cloud computing does present several risks.
It is of paramount importance that client organizations using cloud computing services actively manage such risks (especially security vulnerabilities), regulatory compliance burdens, and the operational inflexibility imposed by vendor lock-in situations.
This paper shows that future advancements in the domain of cloud computing may be optimizing hybrid and multi-cloud architectures for resilience, enhanced edge-cloud continuum for latency-sensitive tasks, and a growing accountability toward sustainable computing.

\vspace{12pt}

\end{document}